\documentclass[12pt]{article}
\pdfoutput=1
\usepackage{epsfig,amsfonts,amsthm}
\usepackage{amsmath,amssymb}
\usepackage[normalem]{ulem}
\usepackage{color}
\usepackage{array}
\newcommand{\be}{\begin{equation}}
\newcommand{\ee}{\end{equation}}
\newcommand{\bea}{\begin{eqnarray}}
\newcommand{\eea}{\end{eqnarray}}

\def\lsim{\mathrel{\rlap{\lower4pt\hbox{\hskip1pt$\sim$}}
    \raise1pt\hbox{$<$}}}         
\def\gsim{\mathrel{\rlap{\lower4pt\hbox{\hskip1pt$\sim$}}
    \raise1pt\hbox{$>$}}}         
    
\usepackage{amsmath}
\usepackage{amsfonts}
\usepackage{amssymb}
\usepackage{subfig}
\usepackage{wrapfig}
\usepackage{graphicx}
\usepackage[normalem]{ulem}

\definecolor{grey}{cmyk}{0,0,0,0.75}
\definecolor{tangerine}{cmyk}{0,0.5,1,0}
\definecolor{darkgreen}{cmyk}{1,0,1,0.23} 
\definecolor{Red}{rgb}{1,0,0}
\definecolor{Blue}{rgb}{0,0,1}
\definecolor{Green}{rgb}{0,1,0}
\definecolor{Grey}{cmyk}{0,0,0,0.75}
\definecolor{Tangerine}{cmyk}{0,0.5,1,0}
\definecolor{Darkgreen}{cmyk}{1,0,1,0.23}
\definecolor{Cyan}{cmyk}{1,0,0,0}
\definecolor{Yellow}{cmyk}{0,0,1,0}

\def\beq{\begin{equation}}
\def\eeq{\end{equation}}
\def\bea{\begin{eqnarray}}
\def\eea{\end{eqnarray}}

\def\<{\left\langle}
\def\>{\right\rangle}

\newcommand{\bt}{\begin{tabular}}
\newcommand{\et}{\end{tabular}}

\usepackage{slashed}
\usepackage{float}
\usepackage[justification=centering]{caption}
\usepackage{tikz}
\usetikzlibrary{decorations.pathmorphing,decorations.markings}
\usepackage{graphicx}
\tikzset{
photon/.style={decorate, decoration={snake,amplitude=2pt, segment length=5pt}, draw=black},
particle/.style={draw=black, postaction={decorate}, decoration={markings,mark=at position .5 with {\arrow[draw=black]{>}}}},
antiparticle/.style={draw=black, postaction={decorate}, decoration={markings,mark=at position .5 with {\arrow[draw=black]{>}}}},
gluon/.style={decorate, draw=black, decoration={coil,amplitude=4pt, segment length=5pt}}
goldstone/.style={draw=green,postaction={decorate},decoration={markings,mark=at position .5 with {\arrow[draw=blue]{>}}}}
}

\usepackage[
top    = 3.5cm,
bottom = 3.5cm,
left   = 3.5cm,
right  = 3.5cm]{geometry}

\allowdisplaybreaks[2]
\begin{document}
\bibliographystyle{OurBibTeX}

\title{
\hfill ~\\[-30mm]
\begin{footnotesize}
\hspace{60mm}
HIP-2018-39/TH\\
\end{footnotesize}
\vspace{5mm}
\textbf{
Cosmological constraints on light flavons
} }
\date{}
\author{\\[-5mm]
Matti Heikinheimo\footnote{E-mail: {\tt Matti.Heikinheimo@helsinki.fi}},\ 
~Katri Huitu\footnote{E-mail: {\tt Katri.Huitu@helsinki.fi}},\
~Venus Keus\footnote{E-mail: {\tt Venus.Keus@helsinki.fi}},\
~Niko Koivunen\footnote{E-mail: {\tt Niko.Koivunen@helsinki.fi}}
\\ \\
  \emph{\small Department of Physics and Helsinki Institute of Physics,}\\
  \emph{\small Gustaf H{\"a}llstr{\"o}min katu 2, FIN-00014 University of Helsinki, Finland}\\[4mm]}
\maketitle

\vspace*{-0.250truecm}
\begin{abstract}
\noindent
{The Froggatt-Nielsen mechanism is a well-motivated framework for generating the fermion mass hierarchy. This mechanism introduces flavons, complex scalars  which are singlet under the Standard Model gauge symmetry and charged under a new global family symmetry.
We make use of a leptophilic flavon to produce the charged lepton Yukawa matrix. The real part of the flavon mixes with the Higgs boson and introduces lepton flavour violating interactions which are bounded by experiment. The imaginary part of the flavon, $\eta$, is a long-lived light particle, whose abundance is restricted by cosmological observations. 
For $m_\eta < 2m_e$ where the decay of $\eta$ to charged leptons is kinematically forbidden,
we identify allowed regions of $m_\eta$ with respect to the vacuum expectation value of the flavon field where all experimental and cosmological constraints are satisfied.
} 
\end{abstract}
 
\thispagestyle{empty}
\vfill
\newpage
\setcounter{page}{1}

\section{Introduction}

The origin of the fermion mass hierarchy is a long-standing problem of the Standard Model (SM). Amongst the many beyond the SM (BSM) scenarios aiming to explain this hierarchy, the Froggatt-Nielsen mechanism \cite{Froggatt:1978nt}  offers a natural solution to this problem. 
This mechanism introduces a spontaneously broken global $U(1)$ symmetry and a new scalar field, the flavon. 
The flavon, $\Phi$, is a singlet under the gauge symmetry of the SM, but charged under the $U(1)$ symmetry along with all other SM particles.
The $U(1)$ symmetry is spontaneously broken as the flavon field acquires a vacuum expectation value (vev), denoted by $v_\phi$. 

The real part of the flavon mixes with the SM Higgs field, resulting in two mass eigenstates.
The lighter state is taken to be the scalar boson with a mass of approximately 125 GeV observed at the Large Hadron Collider (LHC) \cite{Aad:2012tfa,Chatrchyan:2012ufa}.
The imaginary part of the flavon field, $\eta$, is a pseudo-Goldstone boson.

In general, the flavon has flavour violating couplings, which the SM-Higgs boson inherits due to their mixing. As a result, all three scalar mass eigenstates take part in flavour violating interactions. 
The resulting flavour violating processes, which are dominated by $\eta$, 
are severly constrained by experiment, with the most stringent bounds coming from charged lepton flavour violation (CLFV) through the three--body decay  $l_{i}\to l_{j}l_{k}l_{l}$, the $l_{i}\to l_{j}\gamma$ transition and the $\mu \leftrightarrow e$ conversion processes \cite{Bellgardt:1987du}-\cite{Bertl:2006up}.
With the couplings of $\eta$ inversely proportional to $v_\phi$, the non-observation of CLFV processes puts a lower bound on the vev of the flavon, $v_\phi \gtrsim \mathcal{O}$(TeV).

In this paper, we shall study the charged lepton sector and make use of a leptophilic flavon to produce the charged Yukawa mass structure \cite{Huitu:2016pwk,Keus:2017xhw}. Other fermion Yukawa textures could be constructed with the introduction of extra flavons and family symmetries.

We investigate cosmological implications of a very light $\eta$, with $m_\eta<2m_e$ so that its dominant decay channel is into two photons.
The presence of the $\eta$ particles during the early epochs are felt via their contribution to the total energy density of the radiation dominated universe, and via the energy deposited by the decays of $\eta$ to the SM radiation bath during and after the processes of the big bang nucleosynthesis (BBN) and recombination. Thus the model faces severe constraints from the observed abundance of chemical elements \cite{Mangano:2011ar} and the cosmic microwave background (CMB) \cite{Verde:2016wmz,Poulin:2016anj}.
Moreover, with a mass below a few keV, $\eta$ is a hot relic, whose abundance is suppressed by cosmic structure formation at small scales~\cite{Irsic:2017ixq}. 
In the literature, cosmological constraints from BBN and primordial baryon asymmetry have been discussed, with no attention to CMB bounds for a leptophobic flavon \cite{Lillard:2018zts}. Leptophobic flavons have also been considered in the context of electroweak baryogenesis \cite{Baldes:2016gaf}.

The mechanism for production of $\eta$ varies by changing $v_\phi$; for relatively small values, $v_\phi \sim \mathcal{O}(10^4 - 10^8)$ GeV, $\eta$ will be produced as a relativistic relic through a freeze-out mechanism \cite{Fornengo:1997wa} with its abundance well below the observed dark matter (DM) relic density,
\be 
\label{eq:planck}
\Omega_{DM}\,h^2\,=\,0.1197\,\pm\,0.0022 ,
\ee
as measured by the Planck experiment \cite{Ade:2015xua}.
In this range of $v_\phi$, we show that increasing $m_\eta$ leads to a larger initial abundance while reducing the lifetime of $\eta$ considerably, down to $\tau_\eta \sim 10^{10}$~s. Lighter particles, $m_\eta$ of a few meV, have much longer lifetimes, but their abundance is of order $10^{-5}$ with respect to the DM relic density. 
For higher values of the flavon vev, $v_\phi \gtrsim 10^9 $ GeV, the couplings of $\eta$ are so small that it does not thermalise with the SM bath and is produced through a freeze-in mechanism \cite{Hall:2009bx}, with an abundance below that of the DM.

Throughout this paper, we refer to $\Omega_{\eta}h^2$ as the abundance the flavon relic onto which cosmological bounds apply. Note that in regions where $\eta$ is produced through the freeze-out mechanism, relatively light $\eta$s which pass the abundance bounds, have a lifetime longer than the age of the universe. For very heavy $\eta$s ($m_{\eta} \approx 1 $ MeV) the lifetime is so short that no $\eta$ particles are left long enough for any cosmological bounds to be applicable. In the region where the freeze-in mechanism is in play, the couplings of $\eta$ are very small and the lifetime of $\eta$ is much longer than the age of the universe, so in practice $\eta$ could be considered a stable particle.

The layout of the paper is as follows. We first review the Froggatt-Nielsen mechanism in Section \ref{FN-review}, and introduce the scalar potential for the pseudo-Goldstone field in Section \ref{sec: scalar potential}. We discuss CLFV constraints in Section \ref{CLFV-bounds}, and then study the thermal history of the model in Section \ref{thermalisation}. We present the resulting cosmological constraints in Section \ref{sec: cosmo constraints} and draw our conclusions in Section \ref{conclusions}.

\section{The Froggatt-Nielsen mechanism}
\label{FN-review}

The Froggatt-Nielsen mechanism is a well-motivated framework for generating the fermion mass hierarchy \cite{Froggatt:1978nt}. This mechanism introduces a complex scalar field, $\Phi$, called the \emph{flavon}, which is a singlet under the SM gauge group, but charged under a new global $U(1)$ symmetry.
All the SM particles are charged under this global symmetry. 

Consistent with the $U(1)$ charges, the SM Yukawa interactions are generated through higher order operators of the form
\be
\label{FN-operator}
\mathcal{L}\supset c_{ij} ~ \left(\frac{\Phi}{\Lambda}\right)^{n_{ij}} \bar f_{L,i}~ H f_{R,j} + {\rm h.c.} ~,
\ee 
where $c_{ij}$ are dimensionless order-one coefficients, $\Lambda$ is the scale of new-physics, $H$ is the SM Higgs doublet, and $f_{L,R}$ are the SM fermions.
Conservation of the Froggatt-Nielsen symmetry requires the $U(1)$ charges in Eq.(\ref{FN-operator}) to add up to zero, resulting in
\be 
 n_{ij} =-\frac{1}{q_{\Phi}}(q_{\bar L,i}+q_{R,j}+q_{H}) > 0,
\label{FN-powers}
\ee
where $q_{L,R}, q_H, q_\Phi$ are the charges of the SM fermions $f_{R,L}$, the SM Higgs field $H$ and the flavon $\Phi$, respectively.
As the flavon develops a vev, 
$\Phi=(v_\phi+\phi)/ \sqrt{2}$,
the effective operator in Eq.(\ref{FN-operator}) generates the SM Yukawa interactions,
\be
\mathcal{L}_{eff} \supset y_{ij}~ \bar f_{L,i} H f_{R,j} + {\rm h.c.} 
\quad \mbox{with} \quad 
y_{ij} = 
c_{ij} \left(\frac{v_\phi}{\sqrt{2}\Lambda}\right)^{n_{ij}}
\equiv
c_{ij} \epsilon^{n_{ij}},
\ee
where $\epsilon$ is naturally a small parameter ($v_\phi \ll \Lambda$).
Note that the $U(1)$ charge assignment determines the $n_{ij}$ power of $\epsilon$ which in turn determines the Yukawa matrix structure. This is the primary feature of the Froggatt-Nielsen mechanism which relates the fermion mass hierarchy to the $U(1)$ charges of the fermions.
In this paper, we shall only consider the leptonic sector and only allow the leptons and the flavon to transform  under the $U(1)$ symmetry\footnote{Other fermions could be assigned extra flavons and transform under other symmetry groups.}.

When the Higgs field acquires a vev, 
\be
H=\frac{1}{\sqrt{2}}\left(
\begin{array}{c}
0\\
v_h+h
\end{array}
\right),
\ee
the Yukawa Lagrangian becomes
\be
\mathcal{L}_{eff}\supset \frac{y_{ij}}{\sqrt{2}}
\left(v_h+ \; h + \;\frac{v_h}{v_\phi} n_{ij} \; \phi
 \right)\bar{l'}_{L,i}l'_{R,j} + {\rm h.c.}\ ,
\ee
where the primed leptons, $l'$, are the gauge eigenstates. They are related to the mass eigenstates, $l$, by the unitary transformation $U_{L,R}$, 
\be
y_{{\rm diag}}=U_L \; y\; U_R^{\dagger},
\ee
leading to a Yukawa Lagrangian of the form
\be 
\mathcal{L}_{eff}\supset 
\bar{l}_{L} \; \frac{v_h \; y_{{\rm diag}}}{\sqrt{2}} \; l_R + \;
\bar{l}_{L} \; \frac{y_{{\rm diag}}}{\sqrt{2}} \; l_R \; h + \;
\bar{l}_{L} \; \frac{v_h \; \kappa}{\sqrt{2}\; v_\phi}  \;l_R \; \phi + \;
{\rm h.c.}\ .
\ee 
In general, the $\kappa$ matrix,
\be
\kappa=U_L \; (n\cdot y) \; U^{\dagger}_R
\quad \mbox{with} \quad
(n\cdot y)_{ij}=n_{ij} y_{ij},
\ee
is not diagonal and sources the flavour violating processes in the model. In what follows, we parametrise our results in terms of the $\widetilde{\kappa}$ coupling where
\be
\widetilde{\kappa}_{ij}=\frac{1}{\sqrt{2}}\frac{v_h}{v_\phi}\kappa_{ij}. 
\ee
Note that the flavon field is a complex field, explicitly deconstructed as 
\be 
\Phi=\frac{1}{\sqrt{2}}(v_\phi+\phi) \quad \mbox{where} \quad \phi =\sigma + i \; \eta,
\ee
whose leptonic couplings appear as
\be
\mathcal{L}_{l,\phi}=
\bar{l}_{L} \; \widetilde{\kappa} \; l_R \; \sigma
+\bar{l}_{L} \; i \; \widetilde{\kappa} \; l_R  \;\eta + \; {\rm h.c.}\ .
\ee

\section{The scalar potential}
\label{sec: scalar potential}

The Higgs-portal \cite{Silveira:1985rk}-\cite{Patt:2006fw} scalar potential is of the following form,
\bea 
V&=&-\frac{\mu_h^2}{2}(H^\dagger H) 
+ \frac{\lambda_h}{2}(H^\dagger H)^2 
-\frac{\mu_\phi^2}{2}(\Phi^\dagger \Phi) 
+ \frac{\lambda_\phi}{2}(\Phi^\dagger \Phi)^2 \nonumber\\
&&+ \lambda_{h\phi} (H^\dagger H)(\Phi^\dagger \Phi) 
-\frac{{\mu'}^2_\phi}{4}(\Phi^2 +{\Phi^\dagger}^2),
\eea
where the $U(1)$ symmetry is softly broken by the last term, which is responsible for the mass of the $\eta$ field.
The minimisation conditions for the potential are
\be 
\mu_h^2 = v_h^2 \lambda_h + v_\phi^2 \lambda_{h\phi} \;, 
\qquad
\mu_\phi^2 = v_\phi^2 \lambda_\phi + v_h^2 \lambda_{h\phi} - {\mu'}^2_\phi \; .
\ee 
The mass eigenstates $H_1$ and $H_2$ are given by
\be 
\label{mixing}
\left(\begin{array}{c}
H_1\\
H_2
\end{array}\right) \equiv 
\left(\begin{array}{cc}
\cos\theta & -\sin\theta\\
\sin\theta & \cos\theta 
\end{array}\right) \left(\begin{array}{c}
h\\
\sigma 
\end{array}\right) ,
\qquad
\tan 2\theta = \frac{2 \lambda_{h\phi} v_h v_\phi}{\lambda_{h} v_h^2 -\lambda_\phi v_\phi^2} \;.
\ee
The masses of the three physical scalar states are calculated  to be
\be  
m^2_{H_{1,2}}= 
\frac{1}{2} \biggl( \lambda_{h} v_h^2 +\lambda_\phi v_\phi^2  \pm \frac{ \lambda_{h} v_h^2 -\lambda_\phi v_\phi^2}{\cos2\theta} \biggr)
\; , \qquad 
m^2_{\eta} = {\mu'}^2_\phi
\; ,
\ee 
where we take $H_1$ to be the SM-like Higgs boson with $125$ GeV mass.
Note that the mixing of $h$ and $\sigma$, which is constrained by $\sin\theta \lesssim 0.3$ \cite{Ilnicka:2018def}, introduces flavour violating couplings for the $H_1$ state, and allows for the $H_2$ state to couple to SM fermions. 

The Yukawa couplings of the scalar mass eigenstate are explicitly written as
\bea
\label{scalar-Yukawas}
\mathcal{L}_{Y} 
&=&
\bar{l}_L\left(\cos\theta \; \frac{y_{\rm{diag}}}{\sqrt{2}}-\sin\theta~ \widetilde{\kappa}\right)l_R  \; H_1
\\
&+&
\bar{l}_L\left(\sin\theta \; \frac{y_{\rm{diag}}}{\sqrt{2}}+\cos\theta~ \widetilde{\kappa}\right)l_R \; H_2
\nonumber\\
&+&
\bar{l}_L \left(i \; \widetilde{\kappa} \right) l_R \; \eta \; +\rm{h.c.} \; . \nonumber
\eea

In agreement with ref \cite{Ilnicka:2018def}, we take into account all theoretical and experimental bounds applicable to the model. For our analysis, we choose representative values of $m_{H_2}=500$ GeV and $\sin\theta=0.1$.
For $m_{H_2} \gg m_{\eta}$, changing the $H_2$ mass has no tangible effect on the behaviour of the model. $\sin\theta =0.1$ is chosen as a very conservative value to satisfy the experimental bounds. 
Note that at $\sin\theta=0$ the flavon fields (both the real and imaginary components) decouple from the SM, leading to no interesting phenomenology.

Collider signatures of the model are almost identical to those of any singlet extension of the SM, through the Higgs portal \cite{Ilnicka:2018def}. 
Model specific collider signatures of the model should come from the charged lepton couplings of the $\eta$ field. Since the strength of the coupling is proportional to the mass of the lepton, the largest $\eta$ coupling is to the $\tau$ lepton. In principle, $\eta$ could be produced in a lepton collider through the $e^+e^- \to Z^* \to \tau^+\tau^- \eta$ process where $\eta$ is radiating off of the $\tau$ leg. However, the tiny cross section of the process and the decaying $\tau$ final states make the observation of such a process improbable.

The state $\eta$ is a pseudo-Goldstone boson and naturally is assumed to be light. For $m_\eta < 2 m_e$, the decay of $\eta$ to a pair of charged leptons is kinematically forbidden. As a result, $\eta$ decays predominantly to a $\gamma\gamma$ final state through a loop of charged leptons, as shown in Figure \ref{decay-fig}, with the leading order decay amplitude of
\be
|\mathcal{M}_{\eta\to\gamma\gamma}|^2=
\frac{e^4}{32 \pi^4} \; m^4_{\eta}
\biggl(\frac{\tilde{\kappa}_{ee}}{m_e} +\frac{\tilde{\kappa}_{\mu\mu}}{m_\mu} +\frac{\tilde{\kappa}_{\tau\tau}}{m_\tau} \biggr)^2 \;.
\ee 
Therefore, the lifetime of $\eta$ is
\be 
\tau_{\eta}=\frac{1}{\Gamma_{\eta \to \gamma\gamma}} = \frac{32 \pi \; m_\eta}{\lvert\mathcal{M}_{\eta\to\gamma\gamma}\rvert^2} \; ,
\ee
which, depending on $m_\eta$ and $v_\phi$, could be long enough to face severe constraints from cosmological observations, as will be discussed in detail in   the following sections.

\begin{minipage}{\linewidth}
\begin{figure}[H]
\centering
\begin{tikzpicture}[thick,scale=1.0]
\draw (1.5,0) -- node[black,above,xshift=-0.1cm,yshift=0.0cm] {$ $} (1.5,0.03);
\draw[dashed] (0,0) -- node[black,above,xshift=-0.5cm,yshift=0cm] {$\eta$} (1.5,0);
\draw[particle] (1.5,0) -- node[black,above,xshift=0cm,yshift=0cm] {$ $} (3,1);
\draw[decorate,decoration={snake,amplitude=3pt,segment length=10pt}] (4.5,-0.75) -- node[black,above,yshift=0.1cm,xshift=0.0cm] {$\gamma$} (3,-0.75);
\draw[particle](3,-0.75) -- node[black,above,yshift=-0.65cm,xshift=-0.2cm]  {$ $} (1.5,0);
\draw[particle](3,1) -- node[black,above,yshift=-0.4cm,xshift=-0.4cm]  {$l_i$} (3,-0.75);
\draw[decorate,decoration={snake,amplitude=3pt,segment length=10pt}](3,1) -- node[black,above,yshift=0.2cm,xshift=0cm]  {$\gamma$} (4.5,1);
\draw[dashed] (3,-0.75) -- node[black,above,yshift=-0.7cm,xshift=0.0cm] {$ $} (3,-0.78);
\end{tikzpicture}
\caption{The long-lived state $\eta$ with $m_\eta < 2 m_e$ decays predominantly to a $\gamma\gamma$ final state.}
\label{decay-fig} 
\end{figure}
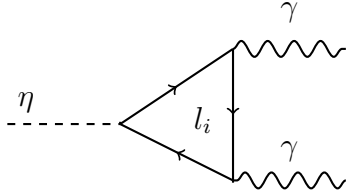    
\end{minipage}

\section{Constraints from CLFV processes}
\label{CLFV-bounds}

We assign the Froggatt-Nielsen charges as shown in Table \ref{charges-table1},
to reproduce the correct charged lepton masses for $\epsilon=0.1$. 
The resulting Yukawa texture and $\widetilde \kappa$ matrix are of the form
\be
\label{rough-texture}
y \sim \left(
\begin{array}{ccc}
 ~\epsilon^6 &  ~\epsilon^5 & ~ \epsilon^4 \\
 ~\epsilon^5 &  ~\epsilon^4  &  ~\epsilon^3 \\
 ~\epsilon^4  &  ~\epsilon^3    &  ~\epsilon^2 \\
\end{array}
\right),
\quad 
\widetilde{\kappa} \sim \frac{1}{v_\phi}\left(
\begin{array}{ccc}
m_e  & m_\mu\;\epsilon  & m_\tau \;\epsilon^2 \\
m_\mu \;\epsilon  & m_\mu  & m_\tau \;\epsilon   \\
m_\tau \;\epsilon^2 & m_\tau \;\epsilon  & m_\tau
\end{array}
\right),
\ee
with the precise values given in Appendix \ref{precise-values}. 

\begin{table}[ht!]
\begin{center}
\begin{tabular}{|c|c|c|c|c|c|c|c|c|}
\hline
Particle   &  $e_L^c$ & $e_R$ & $\mu_L^c$ & $\mu_R$ & $\tau_L^c$ &$\tau_R$ & $H$ & $\Phi$
\\
\hline 
Charge & 3 & 3 & 2 & 2 & 1 & 1 & 0 & -1\\
\hline
\end{tabular}
\end{center}
\vspace{-3mm}
\caption{Our $U(1)$ charge assignment for the SM fields and the flavon. }
\label{charges-table1}
\end{table}

We implement the current bounds from the three--body decay  $l_{i}\to l_{j}l_{k}l_{l}$, the $l_{i}\to l_{j}\gamma$ transition and the $\mu \leftrightarrow e$ conversion processes, presented in Table \ref{experimental-bounds}.  

\begin{table}[h!]
\begin{center}
\begin{tabular}{|p{0.25cm}|l|c|}
\hline  
 & Observable & Present limit \\[1mm]
\hline
1 & BR$(\mu\to eee)$  & $1.0\times 10^{-12}$ \cite{Bellgardt:1987du} 
\\[2mm]
2 & BR$(\tau\to eee)$  & $3.0\times 10^{-8}$ \cite{Amhis:2012bh} 
\\[2mm]
3 & BR$(\tau\to \mu\mu\mu)$  & $2.0\times 10^{-8} $ \cite{Amhis:2012bh}
\\[2mm]
4 & BR$(\tau^{-}\to\mu^{-}e^{+}e^{-})$  & $1.7\times 10^{-8}$ \cite{Hayasaka:2010np}
\\[2mm]
5 & BR$(\tau^{-}\to e^{-}\mu^{+}\mu^{-})$ & $2.7\times 10^{-8}$ \cite{Hayasaka:2010np}
\\[2mm]
6 & BR$(\tau^{-}\to e^{+}\mu^{-}\mu^{-})$ & $1.7\times 10^{-8}$ \cite{Hayasaka:2010np}
\\[2mm]
7 & BR$(\tau^{-}\to \mu^{+}e^{-}e^{-}$) & $1.5\times 10^{-8}$ \cite{Hayasaka:2010np}
\\[2mm]
8 & BR$(\mu\to e\gamma)$ & $4.2\times 10^{-13}$ \cite{Renga:2018fpd}
\\[2mm]
9 & BR$(\tau\to \mu\gamma)$  & $4.4\times 10^{-8}$ \cite{Amhis:2012bh}
\\[2mm]
10 & BR$(\tau\to e\gamma)$ & $3.3\times 10^{-8}$ \cite{Amhis:2012bh}
\\[2mm]
11 & CR$(\mu$-$e, Au)$ & $7.0\times 10^{-13}$ \cite{Bertl:2006up}
\\[1mm]
\hline 
\end{tabular}
\caption{Current experimental bounds on the branching ratios of the three--body decay  $l_{i}\to l_{j}l_{k}l_{l}$, the $l_{i}\to l_{j}\gamma$ transition and the $\mu \leftrightarrow e$ conversion processes.}
\label{experimental-bounds}
\end{center}
\end{table}

All three scalars take part in CLFV interactions. However, with  
$m_\eta < 2 m_e \ll m_{H_1,H_2}$,
the processes mediated by $\eta$ will be by far dominant to those of $H_1$ and $H_2$. Therefore, one can safely disregard these sub-dominant contributions, and only take into account the diagrams shown in Figure \ref{CLFV-processes-fig}.

\begin{minipage}{\linewidth}
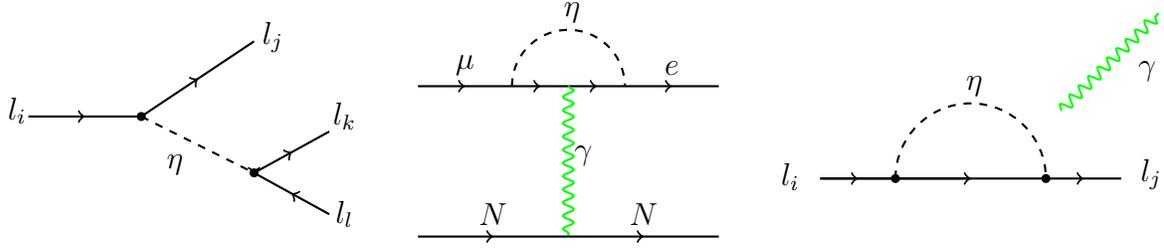
\begin{figure}[H]
\begin{tikzpicture}[thick,scale=1.0]
\fill[black] (1.5,0) circle (0.06cm);
\draw (1.5,0) -- node[black,above,xshift=-0.1cm,yshift=0.0cm] {$ $} (1.5,0.03);
\draw[particle] (0,0) -- node[black,above,xshift=-0.9cm,yshift=-0.25cm] {$l_i$} (1.5,0);
\draw[particle] (1.5,0) -- node[black,above,xshift=1.0cm,yshift=0.2cm] {$l_j$} (3,1.0);
\draw[particle] (3,-0.75) -- node[black,above,yshift=0.15cm,xshift=0.7cm] {$l_k$} (4,-0.2);
\draw[particle] (4,-1.3) -- node[black,above,yshift=-0.6cm,xshift=0.7cm] {$l_l$} (3,-0.75);
\draw[dashed] (1.5,0) -- node[black,above,yshift=-0.55cm,xshift=-0.3cm] {$\eta$} (3,-0.75);
\fill[black] (3,-0.75) circle (0.06cm);
\draw[particle] (3,-0.75) -- node[black,above,yshift=-0.7cm,xshift=0.0cm] {$ $} (3,-0.78);
\end{tikzpicture}
\hspace{4mm}
\begin{tikzpicture}[thick,scale=1.0]
\draw[particle] (0,0) -- node[black,above,sloped,yshift=0.0cm,xshift=0.0cm] {$\mu$} (1.25,0);
\draw[particle] (1.25,0) -- node[black,above,sloped,yshift=0.0cm,xshift=0.0cm] {$ $} (2,0);
\draw[particle] (2,0) -- node[black,above,sloped] {$ $} (2.75,0);
\draw[particle] (2.75,0) -- node[black,above,sloped] {$e$} (4,0);
\draw[decorate,decoration={snake,amplitude=2pt,segment length=5pt},green] (2,0) -- node[black,above,yshift=-0.2cm,xshift=0.2cm] {$\gamma$} (2,-2);
\draw[particle] (0,-2) -- node[black,above,sloped,yshift=0.0cm,xshift=0.0cm] {$N$} (2,-2);
\draw[particle] (2,-2) -- node[black,above,sloped,yshift=0.0cm,xshift=0.0cm] {$N$} (4,-2);
\draw[dashed]  (1.25,0) node[black,above,sloped,yshift=0.7cm,xshift=0.8cm] {$\eta$}  arc (180:0:0.75cm) ;
\end{tikzpicture}
\hspace{4mm}
\begin{tikzpicture}[thick,scale=1.0]
\fill[black] (1,0) circle (0.06cm);
\draw (0,0) -- node[black,above,yshift=-0.8cm,xshift=0.0cm] {$ $} (2,0);
\draw[particle] (0,0) -- node[black,above,sloped,yshift=-0.3cm,xshift=-0.9cm] {$l_i$} (1,0);
\draw[particle] (1,0) -- node[black,above,sloped,yshift=-0.7cm,xshift=0.0cm] {$ $} (3,0);
\draw[particle] (3,0) -- node[black,above,sloped,yshift=-0.3cm,xshift=0.9cm] {$l_j$} (4,0);
\draw[decorate,decoration={snake,amplitude=2pt,segment length=5pt},green] (3.2,0.9) -- node[black,above,yshift=-0.4cm,xshift=0.5cm] {$\gamma$} (4.5,2.2);
\draw[dashed]  (1,0) node[black,above,sloped,yshift=0.95cm,xshift=1.05cm] {$\eta$}  arc (180:0:1cm) ;
\fill[black] (3,0) circle (0.06cm);
\draw (3,0)  node[black,above,yshift=-0.8cm,xshift=0.0cm] {$ $} (3,0);
\end{tikzpicture}
\vspace{4mm}
\caption{The $l_{i}\to l_{j}l_{k}l_{l}$ (left), $\mu\leftrightarrow e$-conversion (center) and $l_{i}\to l_{j}\gamma$ (right) processes mediated by the lightest scalar, $\eta$. In the diagram on the right, the photon can be attached to the internal fermion propagator, or to the external fermion legs. }
\label{CLFV-processes-fig} 
\end{figure}    
\end{minipage}
\vspace{3mm}

The most constraining CLFV limit proves to be the $\mu\to e\gamma$ transition. The branching ratio of this process is
\bea
{\rm BR}(\mu\to e\gamma) 
&=& 
\frac{\alpha \; m_\mu}{4096 \; \pi^4 \; \Gamma^{tot}_{\mu}}
\biggl(\;
\left\lvert
\widetilde{\kappa}_{ee}\widetilde{\kappa}_{e\mu}
+\widetilde{\kappa}^\ast_{\mu e}\widetilde{\kappa}_{\mu\mu}
+\widetilde{\kappa}^\ast_{\tau e}\widetilde{\kappa}_{\mu\tau} \; \frac{m_\mu}{m_\tau}
\right\rvert^2
\nonumber\\
&&
\hspace{25mm}
+\left\lvert
\widetilde{\kappa}_{ee}\widetilde{\kappa}_{\mu e}^\ast
+\widetilde{\kappa}_{e\mu}\widetilde{\kappa}_{\mu\mu}
+\widetilde{\kappa}_{e\tau}\widetilde{\kappa}_{\tau\mu} \; \frac{m_\mu}{m_\tau}
\right\rvert^2 \;
\biggr),
\eea
where we have neglected $m_e$ and $m_\eta$ in comparison to $m_\mu$. Note that this result 
constrains $\widetilde{\kappa}$ (and consequently $v_\phi$) independently of $m_\eta$, since in the studied mass range of $0 < m_\eta < 2 m_e$, the BR($\mu\to e\gamma$) is independent of $m_\eta$.
For our specific charge assignment in Table \ref{charges-table1} and the resulting Yukawa texture in Eq.(\ref{rough-texture}), we obtain 
\be 
{\rm BR}(\mu\to e\gamma) < 4.2 \times 10^{-13} 
\quad \Rightarrow \quad
v_\phi > 14.4 \;{\rm TeV} \;.
\ee

\section{Production of $\eta$ in the early universe}
\label{thermalisation}

For a given mass, the determining factor
in the thermalisation and abundance of $\eta$ is the flavon vev, since the interaction strength of $\eta$ with the SM particles is inversely proportional to $v_\phi$. A relatively small $v_\phi$, of order $ 10^{4} - 10^{8}$ GeV, results in the thermal freeze-out of $\eta$. In this vev range, increasing $v_\phi$ leads to an earlier freeze-out. If the flavon vev is very large, $v_\phi \gtrsim 10^{9}$ GeV, $\eta$ never comes into thermal equilibrium with the SM bath, and is produced through the freeze-in mechanism. 

In the freeze-out scenario, the $\eta$ field is kept in thermal equilibrium with the SM plasma in the early universe primarily 
through the processes shown in Figure \ref{diagrams} and their corresponding $u$- and $t$- channel counterparts.

\begin{minipage}{\linewidth}
\begin{figure}[H]
\begin{center}
\begin{tikzpicture}[thick,scale=1.0]
\draw[decorate,decoration={snake,amplitude=2pt,segment length=5pt},green] (0,0) -- node[black,above,yshift=-0.0cm,xshift=-0.0cm] {$\gamma$} (1,-1);
\draw[particle] (0,-2) -- node[black,above,yshift=0.0cm,xshift=0.0cm] {$l$} (1,-1);
\draw[particle] (1,-1) -- node[black,above,yshift=0.0cm,xshift=0.0cm] {$l$} (2.5,-1);
\draw[dashed] (2.5,-1) -- node[black,above,yshift=0.0cm,xshift=0.0cm] {$\eta$} (3.5,0);
\draw[particle] (2.5,-1) -- node[black,above,yshift=0.0cm,xshift=0.0cm] {$l$} (3.5,-2);
\end{tikzpicture}
\hspace{8mm}
\begin{tikzpicture}[thick,scale=1.0]
\draw[dashed] (0,0) -- node[black,above,yshift=-0.0cm,xshift=0.2cm] {$H_{1,2}$} (1,-1);
\draw[dashed] (0,-2) -- node[black,above,yshift=-0.7cm,xshift=0.3cm] {$H_{1,2}$} (1,-1);
\draw[dashed] (1,-1) -- node[black,above,yshift=0.0cm,xshift=0.0cm] {$H_{1,2}$} (2.5,-1);
\draw[dashed] (2.5,-1) -- node[black,above,yshift=0.0cm,xshift=0.0cm] {$\eta$} (3.5,0);
\draw[dashed] (2.5,-1) -- node[black,above,yshift=0.0cm,xshift=0.0cm] {$\eta$} (3.5,-2);
\end{tikzpicture}
\hspace{8mm}
\begin{tikzpicture}[thick,scale=1.0]
\draw[decorate,decoration={snake,amplitude=2pt,segment length=5pt},black] (0,0) -- node[black,above,yshift=-0.0cm,xshift=0.3cm] {$W,Z$} (1,-1);
\draw[decorate,decoration={snake,amplitude=2pt,segment length=5pt},black] (0,-2) -- node[black,above,yshift=-0.7cm,xshift=0.4cm] {$W,Z$} (1,-1);
\draw[dashed] (1,-1) -- node[black,above,sloped,yshift=0.0cm,xshift=0.0cm] {$H_{1,2}$} (2.5,-1);
\draw[dashed] (2.5,-1) -- node[black,above,yshift=0.0cm,xshift=0.0cm] {$\eta$} (3.5,0);
\draw[dashed] (2.5,-1) -- node[black,above,yshift=0.0cm,xshift=0.0cm] {$\eta$} (3.5,-2);
\end{tikzpicture}
\vspace{3mm}
\caption{The most important number changing diagrams for $\eta$ in the early universe.}
\label{diagrams}
\end{center}
\end{figure}
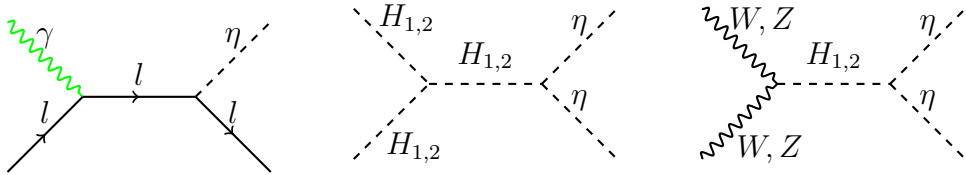
\end{minipage}

We estimate the freeze-out temperature by comparing the Hubble rate, $H$, to the interaction rate $\Gamma$ of the $\eta$-production process $\alpha x\to\eta y$, where
\be
\Gamma_{\alpha x\to\eta y}(T)=n_\alpha\langle v\sigma_{\alpha x\to\eta y} \rangle.
\ee
Here, $\alpha$ represents the heaviest particle in the process (other than $\eta$) which is in thermal  equilibrium with the SM heat bath, and $x$ and $y$ represent other particles involved, e.g. photons. The relativistic number density of the particle species $\alpha$ is
\be
n_\alpha=\left\{\begin{array}{c}
\frac{3}{4\pi^2}\zeta(3)g_\alpha T^3\quad \rm{fermions},\\[2mm]
\frac{1}{\pi^2}\zeta(3)g_\alpha T^3\quad \rm{bosons},\\
\end{array}
\right.
\ee
where $\zeta(3)=1.20206$ is the Riemann zeta function of 3, $g_\alpha$ is the number of  degrees of freedom of particle $\alpha$ and $T$ is the temperature.

When $T\gg m_\alpha$, the thermally averaged cross section scales as $\langle v\sigma_{\alpha x\to\eta y} \rangle\sim 1/s\sim T^{-2}$. Hence the interaction rate $n\langle \sigma v\rangle$ scales as $T$, while the Hubble rate scales as $T^2$ as a function of the temperature. Therefore, in this relativistic regime, the interaction rate becomes faster compared to the Hubble rate as the universe expands and the temperature decreases.

When $\alpha$ becomes non-relativistic at $ T < m_\alpha$, the number density and thus the interaction rate will be exponentially suppressed. Hence, if the process $\alpha x\to\eta y$ is not in equilibrium at $T=m_\alpha$, it never was and it never will be. The following condition can thus be used as a rule of thumb in determining whether the process $\alpha x\to\eta y$ thermalises or not:
\be
\Gamma_{\alpha x\to\eta y}(T=m_\alpha)<H(T=m_\alpha)\Longrightarrow \textrm{no  thermalisation.}
\ee

\begin{figure}[ht!]
\begin{center}
\includegraphics[scale=1.2]{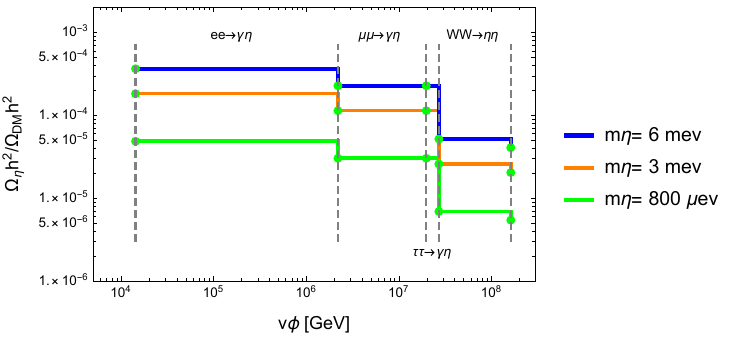}
\caption{Abundance of $\eta$ with respect to $v_\phi$, produced through the freeze-out process for different values of $m_\eta$. The gray dashed vertical lines indicate the $v_\phi$ intervals and the corresponding main production process.}
\label{vOPlot}
\end{center}
\end{figure}

Therefore, the freeze-out temperature is the temperature at which the last number-changing process involving $\eta$, falls out of equilibrium with the SM heat bath which we estimate to be $T_{fo}\sim m_\alpha$.
This treatment results in a step-like abundance of $\eta$ as a function of the flavon vev. 
In Figure \ref{vOPlot}, we show this behaviour for different $\eta$ masses, and identify the thermalisation process in each $v_\phi$ interval. For higher vevs, $\eta$ does not thermalise with the SM heat bath and is produced through a freeze-in mechanism. Note that the lower bound on $v_\phi$ is imposed by the CLFV constraints.

In calculating the thermally averaged cross section, we use the 
standard approximation~\cite{Fornengo:1997wa}  where instead of integrating over the exact statistics of relativistic particles, we assume head-on collisions and approximate the initial energies of the incoming particles with their average thermal energies, given by
\be
\label{thermal energies}
\langle E\rangle =\left\{\begin{array}{c}
\frac{7\pi^4}{180\zeta(3)}T\approx 3.151 T\quad \rm{fermions},
\\[2mm]
\frac{\pi^4}{30\zeta(3)}T\approx 2.701 T\quad \rm{bosons}.\\
\end{array}
\right.
\ee
The yield of the relativistic particle $\eta$, produced in a freeze-out process, at present-day is
\be
Y_{\eta}=\frac{45}{2\pi^4}\zeta (3)\frac{g_{\rm eff,\eta}}{g_{\ast s}(x_f)},
\ee
where $g_{\rm eff,\eta}=1$ is the effective degrees of freedom of $\eta$, $g_{*s}$ is the number of relativistic degrees of freedom related to entropy density and $x_f=m_\eta/T_{fo}$ where the $T_{fo}$ is the freeze-out temperature of particle $\eta$.
The yield is related to the abundance through the relation
\be
\Omega_{\eta}h^2=\frac{m_\eta s_0 Y_{\eta}}{\left(\frac{3H_0^2}{8\pi G_N}\right)}h^2,
\ee
where $s_0=2970 \; {cm}^{-3}$ is the entropy density today, $H_0=(100 \mbox{ KM/s/MPc})h$ is the Hubble rate today with the dimensionless Hubble parameter $h=0.7$ and $G_N$ is the gravitational constant, $G_N=1/m_{PL}^2$ where $m_{PL}$ is the non-reduced Planck mass.

Here, $\Omega_{\eta}h^2$ is the abundance of the flavon relic onto which cosmological bounds apply. Note that in regions where $\eta$ is produced through the freeze-out mechanism, relatively light $\eta$s which pass the abundance bounds, have a lifetime longer than the age of the universe. For very heavy $\eta$s ($m_{\eta} \approx 1 $ MeV) the lifetime is so short that no $\eta$ particles are left long enough for any cosmological bounds to be applicable.

On the other hand, if the flavon vev is very large, which leads to very small $\eta$-SM couplings, the $\eta$ field will never come into thermal equilibrium with the SM bath. In this scenario the $\eta$-abundance is produced via the freeze-in mechanism \cite{Hall:2009bx} due to $AB\rightarrow \eta\eta$ and $AB\rightarrow \eta C$ processes, where $A,B,C$ refer to SM particles. For most of the parameter space in the freeze-in regime, the dominant production channel is the $e e\rightarrow \gamma\eta $ process. The cross sections for the relevant processes are given in the Appendix \ref{detailed-calculations}, and the solution of the Boltzmann equation for the freeze-in production in Appendix \ref{freeze-in}.

In the region where the freeze-in mechanism is in play, the couplings of $\eta$ are very small and the lifetime of $\eta$ is much longer than the age of the universe, so in practice $\eta$ could be considered a stable particle.


To clarify the discussion above, we treat the case of $m_\eta = 1 $ meV, as an instructive example, to outline the thermal history of the field $\eta$ as a function of the flavon vev.
\begin{itemize}
\item 
For values of $v_\phi >  1.622 \times 10^{8}$ GeV, $\eta$ will be produced through the freeze-in mechanism and contributes to a fraction of the DM abundance.
\item 
If $v_\phi \leq  1.622  \times 10^{8}$ GeV, $\eta$ reaches thermal equilibrium with the SM particles. The freeze-out temperature reduces with decreasing $v_\phi$, as lighter SM particles are able to remain in equilibrium with $\eta$.
For $v_\phi > 3.16 \times 10^{7}$ GeV, the freeze-out temperature is $T_{fo}\sim m_{W}$, 
for $v_\phi >2.25 \times 10^{7}$ GeV it is $T_{fo}\sim m_\tau$, 
and for $v_\phi > 1.85 \times 10^{6}$ GeV it is $T_{fo}\sim m_\mu$. 
Below this value, $\eta$ remains in equilibrium during BBN, and therefore contributes to the effective number of neutrinos $N_{\rm eff}$, where the constraint $\Delta N_{\rm eff} < 1$ \cite{Mangano:2011ar} marginally allows the presence of one scalar degree of freedom. $\eta$ will then decouple at the temperature $T_{fo}\sim m_e$.
\end{itemize}

\begin{figure}[h!!]
\begin{center}
\includegraphics[scale=1.1]{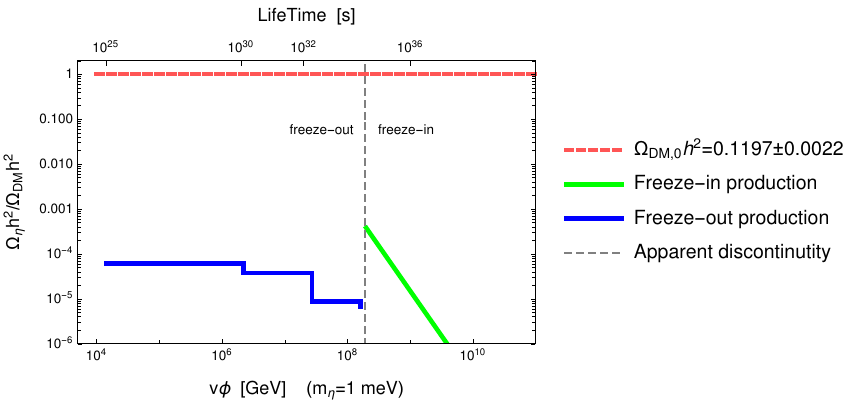}
\caption{Abundance of $\eta$ with respect to the flavon vev and the lifetime of $\eta$ for $m_\eta=1$ meV. The particle $\eta$ is produced through the freeze-out mechanism for $v_\phi \leq  1.622  \times 10^{8}$ GeV, and through the freeze-in process for $v_\phi >  1.622  \times 10^{8}$ GeV.}
\label{FIFO}
\end{center}
\end{figure}

Figure \ref{FIFO} shows the abundance of $\eta$ with respect to $v_\phi$ for $m_\eta = 1$ meV. 
Note that the apparent discontinuity represented by the gray dashed vertical line is due to the assumption that for $v_\phi >  1.622  \times 10^{8}$ GeV, $\eta$ interacts so feebly with the SM bath that it never thermalises. 
As a result, production of $\eta$ goes abruptly from a freeze-out to a freeze-in mechanism. The realistic treatment of this transition should be done by solving the Boltzmann equation numerically which is out of the scope of this paper.
As it will be discussed in the next section, for applying the cosmological constraints,
it is useful to present the abundance of $\eta$ in terms of its lifetime. Hence, for our example of $m_\eta = 1$ meV, we also show the lifetime of $\eta$ corresponding to the values of $v_\phi$ in Figure \ref{FIFO} on the top horizontal axis.

\section{Cosmological constraints}
\label{sec: cosmo constraints}
While $\eta$ is relativistic, it contributes to the total energy density of the radiation dominated universe. 
Additionally, as $\eta$ decays to photons, it deposits energy  to the SM radiation bath during and after the processes of BBN and recombination. The abundance of $\eta$ is therefore constrained from the observed abundance of chemical elements and the CMB \cite{Mangano:2011ar}-\cite{Poulin:2016anj}.
Moreover, for values of $m_\eta \lesssim 3$ keV, $\eta$ is a hot relic and can not constitute the majority of DM, as it would suppress cosmic structure formation at small scales \cite{Irsic:2017ixq}. Therefore, only a small sub-dominant component of hot DM, with a density of $\mathcal{O}(10^{-2})$ of $\Omega_{DM}h^2$ or so, is feasible.

\begin{figure}[h!!]
\begin{center}
\includegraphics[scale=1.1]{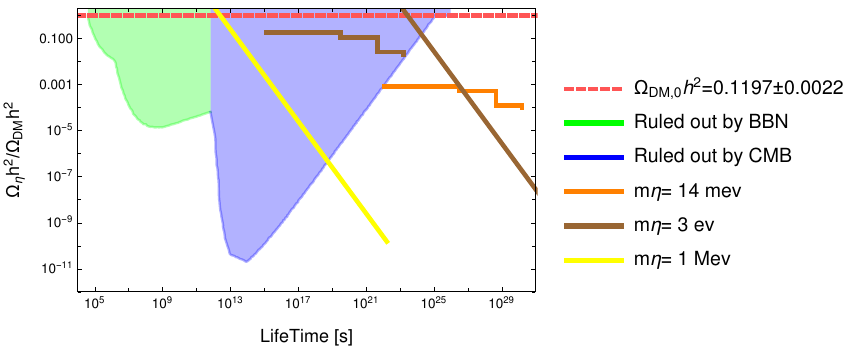}
\caption{The abundance vs. lifetime of $\eta$ for different $m_\eta$. The green and blue areas show the excluded regions from BBN and CMB observations \cite{Poulin:2016anj}, respectively.}
\label{abundance-mass1}
\end{center}
\end{figure}

For a direct comparison with the results of \cite{Poulin:2016anj}, in Figure \ref{abundance-mass1}, we present constraints on the abundance of $\eta$ normalised to the abundance of DM, $\Omega_{\eta}h^2/\Omega_{DM}h^2$, evaluated at present-day, as a function of the lifetime of $\eta$.

The relevant bounds in each mass range are as follows.
For $m_\eta \leq 14$ meV, the abundance of $\eta$ is very low and its lifetime is very long, so no cosmological process imposes any constraints on the parameter space regardless of the freeze-out or freeze-in production of $\eta$. Note that in this mass range, the lower bound of $v_\phi$ is only imposed by the CLFV constraints.

In the intermediate mass range of 14 meV $< m_\eta < 3$  eV, both production mechanisms of $\eta$, freeze-out and freeze-in, are important. 
For such small $m_\eta$ values, $\eta$ is a hot relic and can only contribute a small fraction of the DM density.

For values of 3 eV$ < m_\eta < 2m_e$, freeze-out production of $\eta$ leads to large abundance and short lifetime, ruled out by the CMB data. However, the freeze-in production of $\eta$ allows for very small densities of $\eta$ to survive the cosmological bounds.

In Figure \ref{abundance-mass1}, we show the abundance of $\eta$ for $m_\eta=14$ meV, in orange, where all CLFV-allowed values of $v_\phi > 14.4$ TeV, survive the cosmological constraints. 
The abundance of $m_\eta = 3$ eV is shown in brown where all $v_\phi$ values that lead to a freeze-out production of $\eta$ are ruled out, but $v_\phi$ values leading to the freeze-in production of $\eta$ are allowed, provided the $\eta$ abundance is below the CMB bounds and is less than 1\% of DM relic density, which we take as an estimate for the allowed abundance of hot DM. A more detailed analysis on the allowed hot relic abundance would require numerical simulations of small scale structure formation with both hot and cold DM components, and is beyond the scope of this work.
The abundance of $m_\eta=1$ MeV $<2m_e$ is shown in yellow where only very small densities of $\eta$ are allowed for very large values of $v_\phi$. The graph only shows $v_\phi$ values up to $10^{16}$ GeV. 

\begin{figure}[h!]
\begin{center}
\includegraphics[scale=1.1]{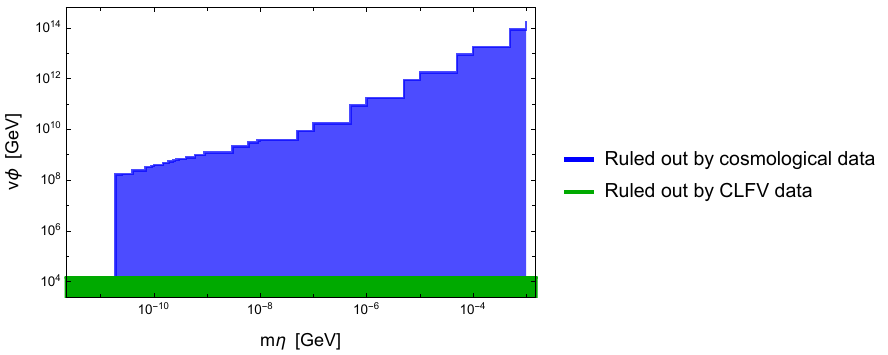}
\caption{Regions ruled out in the $m_\eta$-$v_\phi$ plane, due to the constraints from 
CMB data and small scale structure observations, in blue, and from 
CLFV experiments, in green.}
\label{final-plot}
\end{center}
\end{figure}

In Figure \ref{final-plot}, we show the allowed region in the  $m_\eta$-$v_\phi$ plane. As mentioned before, the lower bound on the flavon vev comes from the CLFV experiments, irrespective of $m_\eta$, represented by the green area. 
The blue area represents the region where the abundance of $\eta$ is constrained by CMB data and small scale structure observations.

\section{Conclusions}
\label{conclusions}
 
In this paper, we have used the Froggatt-Nielsen mechanism to generate the charged lepton Yukawa matrix with a leptophilic flavon whose real part couples to the SM Higgs field and its imaginary part, $\eta$, is a light pseudo-Goldstone boson.
The resulting flavour violating couplings are constrained by the non-observation of the CLFV processes which put a lower bound on the vev of the flavon.

The production mechanism for $\eta$ is determined by the value of the flavon vev where for relatively small values of $v_\phi$ of order $10^4 - 10^8$ GeV, $\eta$ is produced relativistically through a freeze-out mechanism
with its abundance below the observed DM relic density.
In this range of $v_\phi$, we show that increasing $m_\eta$ leads to a larger abundance while reducing its lifetime.
Smaller $\eta$ masses, below the meV range, lead to longer lifetimes and much smaller relic densities.
For high values of $v_\phi$, $\eta$ will be produced through a freeze-in mechanism.

We study cosmological implications of such light $\eta$ particles with $m_\eta<2m_e$, whose dominant decay channel is into two photons. 
In this mass range, we show that the abundance of the $\eta$ is limited by BBN, CMB and small scale structure observations, and identify the allowed region in the $m_\eta$-$v_\phi$ space.


\subsection*{Acknowledgements}  

KH and VK acknowledge the H2020-MSCA-RISE-2014 grant no. 645722 (NonMinimalHiggs). NK is supported by Vilho, Yrj{\"o} and Kalle V{\"a}is{\"a}l{\"a} Foundation.


\appendix

\section{The Yukawa texture}
\label{precise-values}
The charged lepton Yukawa matrix with the order-one coefficients is
\be
\label{texture}
y= \left(
\begin{array}{ccc}
3.812~ \epsilon^6 & -0.625~ \epsilon^5 & 3.5~\epsilon^4 \\
1.36 ~\epsilon^5 &  5.624 ~\epsilon^4  &  -0.7 ~\epsilon^3 \\
0.5~ \epsilon^4  &  0.7~\epsilon^3    &  1.0147 ~\epsilon^2 \\
\end{array}
\right),
\ee
with the $\widetilde{\kappa}$ matrix as
\be
\label{kappa}
\widetilde{\kappa} = \frac{\rm{GeV}}{v_\phi}\left(
\begin{array}{ccc}
0.00305512  & 0.00478025  & 0.115462  \\
-0.00285967  & 0.423056 & 0.11364 \\
0.0139282 & -0.108035 & 3.56724
\end{array}
\right).
\ee

\section{Thermalisation of $\eta$}
\label{detailed-calculations}

\subsubsection*{The $H_{1,2} H_{1,2}\to\eta\eta$ process}
The diagrams contributing to the process $H_{1,2} H_{1,2}\to\eta\eta$ are shown in Figure \ref{HH-diags}.

\begin{minipage}{\linewidth}
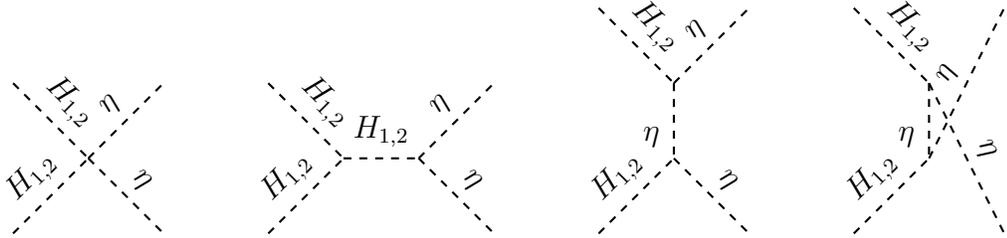
\begin{figure}[H]
\begin{center}
\begin{tikzpicture}[thick,scale=1.0]
\draw[dashed] (0,0) -- node[black,above,sloped,yshift=-0.0cm,xshift=-0.0cm] {$H_{1,2}$} (1,-1);
\draw[dashed] (1,-1) -- node[black,above,sloped,yshift=0.0cm,xshift=0.0cm] {$\eta$} (2,0);
\draw[dashed] (0,-2) -- node[black,above,sloped,yshift=0.0cm,xshift=0.0cm] {$H_{1,2}$} (1,-1);
\draw[dashed] (1,-1) -- node[black,above,sloped,yshift=0.0cm,xshift=0.0cm] {$\eta$} (2,-2);
\end{tikzpicture}
\hspace{7mm}
\begin{tikzpicture}[thick,scale=1.0]
\draw[dashed] (0,0) -- node[black,above,sloped,yshift=-0.0cm,xshift=-0.0cm] {$H_{1,2}$} (1,-1);
\draw[dashed] (0,-2) -- node[black,above,sloped,yshift=0.0cm,xshift=0.0cm] {$H_{1,2}$} (1,-1);
\draw[dashed] (1,-1) -- node[black,above,sloped,yshift=0.0cm,xshift=0.0cm] {$H_{1,2}$} (2,-1);
\draw[dashed] (2,-1) -- node[black,above,sloped,yshift=0.0cm,xshift=0.0cm] {$\eta$} (3,0);
\draw[dashed] (2,-1) -- node[black,above,sloped,yshift=0.0cm,xshift=0.0cm] {$\eta$} (3,-2);
\end{tikzpicture}
\hspace{7mm}
\begin{tikzpicture}[thick,scale=1.0]
\draw[dashed] (0,0) -- node[black,above,sloped,yshift=-0.0cm,xshift=-0.0cm] {$H_{1,2}$} (1,-1);
\draw[dashed] (1,-1) -- node[black,above,sloped,yshift=0.0cm,xshift=0.0cm] {$\eta$} (2,0);
\draw[dashed] (1,-1) -- node[black,above,yshift=-0.5cm,xshift=-0.3cm] {$\eta$} (1,-2);
\draw[dashed] (0,-3) -- node[black,above,sloped,yshift=0.0cm,xshift=0.0cm] {$H_{1,2}$} (1,-2);
\draw[dashed] (1,-2) -- node[black,above,sloped,yshift=0.0cm,xshift=0.0cm] {$\eta$} (2,-3);
\end{tikzpicture}
\hspace{7mm}
\begin{tikzpicture}[thick,scale=1.0]
\draw[dashed] (0,0) -- node[black,above,sloped,yshift=-0.0cm,xshift=-0.0cm] {$H_{1,2}$} (1,-1);
\draw[dashed] (1,-1) -- node[black,above,sloped,yshift=0.0cm,xshift=0.0cm] {$\eta$}(2,-3);
\draw[dashed] (1,-1) -- node[black,above,yshift=-0.5cm,xshift=-0.3cm] {$\eta$} (1,-2);
\draw[dashed] (0,-3) -- node[black,above,sloped,yshift=0.0cm,xshift=0.0cm] {$H_{1,2}$} (1,-2);
\draw[dashed] (1,-2) -- node[black,above,sloped,yshift=0.0cm,xshift=0.0cm] {$\eta$} (2,0) ;
\end{tikzpicture}
\vspace{2mm}
\caption{The diagrams contributing to the $H_{1,2} H_{1,2}\to\eta\eta$ process.}
\label{HH-diags}
\end{center} 
\end{figure}    
\end{minipage} 
\vspace{3mm}

With $m_\eta <2m_e \ll m_{H_1,H_2}$, the $H_1 H_1\to\eta\eta$ cross section is calculated to be
\bea
&&
\sigma_{H_1 H_1\to \eta\eta}=
\\
&&
\frac{1}{64\pi E_{AH_1} E_{BH_1}}\frac{1}{v_{rel}}\Bigg\{ 2\left[(2!)(2!)\lambda_{\eta\eta 11}
-\frac{(2!)\lambda_{\eta\eta 1}(3!)\lambda_{111}}{s-m_{1}^2}
-\frac{(2!)\lambda_{\eta\eta 2}(2!)\lambda_{211}}{s-m_{2}^2}\right]^2
\nonumber\\ 
&&
+ 2 \left[(2!)(2!)\lambda_{\eta\eta 11}
-\frac{(2!)\lambda_{\eta\eta 1}(3!)\lambda_{111}}{s-m_{1}^2}
-\frac{(2!)\lambda_{\eta\eta 2}(2!)\lambda_{211}}{s-m_{2}^2}\right] \frac{16\lambda_{\eta\eta 1}^2 }{s\;\beta_{1}}\ln\left[\frac{1-2k_{1}+\beta_{1}}{1-2k_{1}-\beta_{1}}\right]\nonumber\\
&& 
+ (2!)^4\frac{8\lambda_{\eta\eta 1}^4}{s^2}\left[-\frac{2}{\beta^2_{1}-(1-2k_{1})^2}
+\frac{1}{\beta_{1}(1-2k_{1})}\ln\left(\frac{1-2k_{1}+\beta_{1}}{1-2k_{1}-\beta_{1}}\right)\right]\Bigg\}\frac{1}{2!}, \nonumber
\eea
where we have used the following notation
\be 
\lambda_{ijk}=\lambda_{H_iH_jH_k},
\qquad
\lambda_{\eta\eta i}=\lambda_{\eta \eta H_i},
\qquad
\lambda_{\eta\eta i j}=\lambda_{\eta \eta H_i H_j},
\ee
and
\be
k_i=\frac{m_{H_i}^2}{s}, 
\qquad \beta_i=\sqrt{1-\frac{4m_{H_i}^2}{s}},
\qquad
m_i=m_{H_i},
\ee
and $E_{AH1}$ and $E_{BH1}$ are the initial energies of the scalars.

Similarly, the cross section for the $H_2 H_2\to\eta\eta$ is calculated to be
\bea
&&
\sigma_{H_2 H_2\to \eta\eta}=
\\
&&
\frac{1}{64\pi E_{AH_2} E_{BH_2}}\frac{1}{v_{rel}}\Bigg\{ 2\left[(2!)(2!)\lambda_{\eta\eta 22}
-\frac{(2!)\lambda_{\eta\eta 1}(2!)\lambda_{122}}{s-m_{1}^2}
-\frac{(2!)\lambda_{\eta\eta 2}(3!)\lambda_{222}}{s-m_{2}^2}\right]^2
\nonumber\\ 
&&
+ 2 \left[(2!)(2!)\lambda_{\eta\eta 22}
-\frac{(2!)\lambda_{\eta\eta 1}(2!)\lambda_{122}}{s-m_{1}^2}
-\frac{(2!)\lambda_{\eta\eta 2}(3!)\lambda_{222}}{s-m_{2}^2}\right] \frac{16\lambda_{\eta\eta 2}^2 }{s \;\beta_{2}}\ln\left[\frac{1-2k_{2}+\beta_{2}}{1-2k_{2}-\beta_{2}}\right]
\nonumber\\
&& 
+ (2!)^4\frac{8\lambda_{\eta\eta 2}^4}{s^2}\left[-\frac{2}{\beta^2_{2}-(1-2k_{2})^2}
+\frac{1}{\beta_{2}(1-2k_{2})}\ln\left(\frac{1-2k_{2}+\beta_{2}}{1-2k_{2}-\beta_{2}}\right)\right]\Bigg\}\frac{1}{2!}, \nonumber
\eea
and for the $H_1 H_2\to\eta\eta$ process to be
\bea
&&
\sigma_{H_1 H_2\to \eta\eta}=
\\
&&
\frac{1}{64\pi E_{AH_1} E_{BH_2}}\frac{1}{v_{rel}}\Bigg\{  2\left[(2!)(1!)\lambda_{\eta\eta 12}
-\frac{(2!)\lambda_{\eta\eta 1}(2!)\lambda_{112}}{s-m_{1}^2}
-\frac{(2!)\lambda_{\eta\eta 2}(2!)\lambda_{221}}{s-m_{2}^2}\right]^2
\nonumber\\ 
&&
+ 2 \left[(2!)(1!)\lambda_{\eta\eta 12}
-\frac{(2!)\lambda_{\eta\eta 1}(2!)\lambda_{112}}{s-m_{1}^2}
-\frac{(2!)\lambda_{\eta\eta 2}(2!)\lambda_{221}}{s-m_{2}^2}\right]
 \frac{16\lambda_{\eta\eta 1}\lambda_{\eta\eta 2} }{s\beta_{12}}\ln\left[\frac{1-k_{1}-k_{2}+\beta_{12}}{1-k_{1}-k_{2}-\beta_{12}}\right]
\nonumber\\
&& 
+ (2!)^4\frac{8\lambda_{\eta\eta 1}^2\lambda_{\eta\eta 2}^2}{s^2}\left[\frac{-2}{\beta^2_{12}-(1-k_{1}-k_{2})^2}
+\frac{1}{\beta_{12}(1-k_{1}-k_{2})}\ln\left(\frac{1-k_{1}-k_{2}+\beta_{12}}{1-k_{1}-k_{2}-\beta_{12}}\right)\right]\Bigg\}\frac{1}{2!}, 
\nonumber
\eea
where
\be
\beta_{ij}=\sqrt{1-\frac{2(m_{H_i}^2+m_{H_j}^2)}{s}+\frac{(m_{H_i}^2-m_{H_j}^2)^2}{s^2}}.
\ee

\subsubsection*{The $\gamma l_i\to\eta l_i$ process}
The process $\gamma l_i\to\eta l_i$ proceeds through the diagrams presented in Figure \ref{leptonic production of eta b}.

\begin{minipage}{\linewidth}
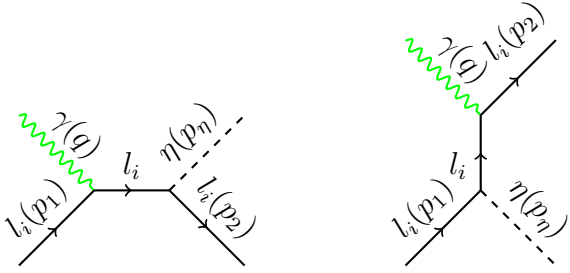
\begin{figure}[H]\begin{center}
\begin{tikzpicture}[thick,scale=1.0]
\draw[decorate,decoration={snake,amplitude=2pt,segment length=5pt},green] (0,0) -- node[black,above,sloped,yshift=-0.0cm,xshift=-0.0cm] {$\gamma(q)$} (1,-1);
\draw[particle] (0,-2) -- node[black,above,sloped,yshift=0.0cm,xshift=0.0cm] {$l_i(p_1)$} (1,-1);
\draw[particle] (1,-1) -- node[black,above,sloped,yshift=0.0cm,xshift=0.0cm] {$l_i$} (2,-1);
\draw[dashed] (2,-1) -- node[black,above,sloped,yshift=0.0cm,xshift=0.0cm] {$\eta(p_\eta)$} (3,0);
\draw[particle] (2,-1) -- node[black,above,sloped,yshift=0.0cm,xshift=0.0cm] {$l_i(p_2)$} (3,-2);
\end{tikzpicture}
\hspace{10mm}
\begin{tikzpicture}[thick,scale=1.0]
\draw[decorate,decoration={snake,amplitude=2pt,segment length=5pt},green] (0,0) -- node[black,above,sloped,yshift=-0.0cm,xshift=-0.0cm] {$\gamma(q)$} (1,-1);
\draw[particle] (1,-1) -- node[black,above,sloped,yshift=0.1cm,xshift=0.4cm] {$l_i(p_2)$} (2,0);
\draw[particle] (1,-2) -- node[black,above,yshift=-0.5cm,xshift=-0.3cm] {$l_i$} (1,-1);
\draw[particle] (0,-3) -- node[black,above,sloped,yshift=0.0cm,xshift=0.0cm] {$l_i(p_1)$} (1,-2);
\draw[dashed] (1,-2) -- node[black,above,sloped,yshift=0.0cm,xshift=0.0cm] {$\eta(p_\eta)$} (2,-3);
\end{tikzpicture}
\caption{Diagrams (a) and (b) contributing to the $\gamma l_i\to\eta l_i$ process.}
\label{leptonic production of eta b}
\end{center}
\end{figure}     
\end{minipage}

The spin averaged amplitude squared for this process is: 
\be 
\lvert\mathcal{M} \rvert^2=\lvert\mathcal{M}_a+\mathcal{M}_b \rvert^2
=\lvert\mathcal{M}_a\rvert^2+\lvert\mathcal{M}_b \rvert^2+
2\mathcal{M}_a^\dagger\mathcal{M}_b, 
\ee
where
\bea
\lvert\mathcal{M}_a\rvert^2
&=&
\frac{(\widetilde{\kappa}_{ii} \; e)^2}{(p_1\cdot q)^2}\left[m_i^2(p_1\cdot q-p_2\cdot q-p_1\cdot p_2+m_i^2)+(p_1\cdot q)(p_2\cdot q)\right],
\nonumber\\
\lvert\mathcal{M}_b\rvert^2
&=&
\frac{(\widetilde{\kappa}_{ii} \; e)^2}{(p_2\cdot q)^2}\left[m_i^2(p_1\cdot q-p_2\cdot q-p_1\cdot p_2+m_i^2)+(p_1\cdot q)(p_2\cdot q)\right],
\nonumber\\
2\mathcal{M}_a^\dagger\mathcal{M}_b
&=&
-2\frac{(\widetilde{\kappa}_{ii} \; e)^2}{(p_1\cdot q)(p_2\cdot q)}\left[m_i^2(m_i^2+p_1\cdot p_2+p_1\cdot q-p_2\cdot q)\right.
\nonumber\\
&&
\left.+(m_i^2+p_1\cdot p_2+p_1\cdot q)(p_1\cdot p_2+p_2\cdot q-m_i^2)\right]
\eea

\subsubsection*{The $l_i l_i\to\gamma\eta$ process}
The process $l_i l_i \to\gamma\eta$ proceeds through the diagrams presented in Figure \ref{leptonic production of eta d}.

\begin{minipage}{\linewidth}
\begin{figure}[H]
\begin{center}
\begin{tikzpicture}[thick,scale=1.0]
\draw[particle] (0,0) -- node[black,above,sloped,yshift=-0.0cm,xshift=-0.0cm] {$l_i (p_1)$} (1,-1);
\draw[decorate,decoration={snake,amplitude=2pt,segment length=5pt},green] (1,-1) -- node[black,above,sloped,yshift=0.1cm,xshift=0.4cm] {$\gamma (q)$} (2,0);
\draw[particle] (1,-1) -- node[black,above,yshift=-0.5cm,xshift=-0.3cm] {$l_i$} (1,-2);
\draw[particle] (1,-2) -- node[black,above,sloped,yshift=0.0cm,xshift=0.0cm] {$l_i(p_2)$} (0,-3);
\draw[dashed] (1,-2) -- node[black,above,sloped,yshift=0.0cm,xshift=0.0cm] {$\eta(k)$} (2,-3);
\end{tikzpicture}
\hspace{10mm}
\begin{tikzpicture}[thick,scale=1.0]
\draw[particle] (0,0) -- node[black,above,sloped,yshift=-0.0cm,xshift=-0.0cm] {$l_i (p_1)$} (1,-1);
\draw[decorate,decoration={snake,amplitude=2pt,segment length=5pt},green] (1,-2) -- node[black,above,sloped,yshift=0.1cm,xshift=0.4cm] {$\gamma (q)$} (2,-0.5);
\draw[particle] (1,-1) -- node[black,above,yshift=-0.5cm,xshift=-0.3cm] {$l_i$} (1,-2);
\draw[particle] (1,-2) -- node[black,above,sloped,yshift=0.0cm,xshift=0.0cm] {$l_i(p_2)$} (0,-3);
\draw[dashed] (1,-1) -- node[black,above,sloped,yshift=0.0cm,xshift=0.0cm] {$\eta(k)$} (2,-2.5);
\end{tikzpicture}
\caption{Diagrams (c) and (d) contributing to the $l_i l_i\to\eta \gamma $ process.}
\label{leptonic production of eta d}
\end{center}
\end{figure}
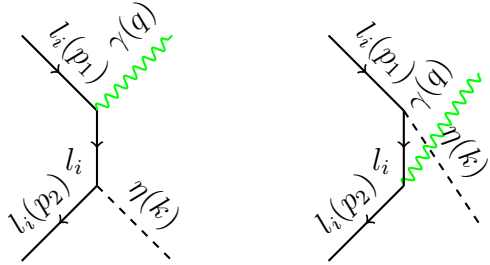
\end{minipage}

The spin averaged amplitude squared for this process is:
\be 
\lvert\mathcal{M} \rvert^2=\lvert\mathcal{M}_c+\mathcal{M}_d \rvert^2
=\lvert\mathcal{M}_c\rvert^2+\lvert\mathcal{M}_d \rvert^2+
2\mathcal{M}_c^\dagger\mathcal{M}_d, 
\ee
where
\bea
\lvert\mathcal{M}_c\rvert^2
&=&
\left[\frac{\widetilde{\kappa}_{ii} \; e}{m_\eta^2-2k\cdot p_2}\right]^2\Big[4(k\cdot p_1)(k\cdot p_2)-2m_\eta^2 (p_1\cdot p_2)-4m_i^2 m_\eta^2\Big],
\nonumber\\
\lvert\mathcal{M}_d\rvert^2
&=&
\left[\frac{\widetilde{\kappa}_{ii} \; e}{m_\eta^2-2k\cdot p_1}\right]^2\Big[4(k\cdot p_1)(k\cdot p_2)-2m_\eta^2 (p_1\cdot p_2)-4m_i^2 m_\eta^2\Big]
\nonumber\\
2\mathcal{M}_c^{\dagger}\mathcal{M}_d
&=&
\frac{(\widetilde{\kappa}_{ii} \; e)^2}{\left[m_\eta^2-2k\cdot p_2\right]\left[m_\eta^2-2k\cdot p_1\right]}\Big[8(k\cdot p_1)(k\cdot p_2)-4m_i^2 m_\eta^2\Big].
\eea

To calculate the thermally averaged cross section, we approximate the initial energies of the photon and the lepton $l_i$ with their average thermal energies, as shown in Eq. (\ref{thermal energies}). Note that the incoming particles have different statistics and therefore different momenta. As a result, the center-of-momentum frame is of no use here. We will calculate the $\gamma l_i\to\eta l_i$ cross section in the general co-linear frame where the incoming particles will collide head-on with non-equal momenta.

\subsubsection*{The $\gamma l_i\to \eta l_i$ cross section in the general co-linear frame}
The cross section for the process $\gamma (\vec{q)} l_i (\vec{p}_1)\to \eta (\vec{p}_\eta) l_i (\vec{p}_2)$ in the general co-linear frame is given by    
\be 
\sigma_{\gamma l_i\to \eta l_i}=\int d\Omega\frac{1}{2E_1 2E_q v_{rel}}\frac{\lvert\vec{p}_2\rvert}{16\pi^2}\left[E_\eta+E_2\left(1-\frac{(\lvert\vec{p}_1\rvert-\lvert\vec{q}\rvert)\cos\omega}{\lvert\vec{p}_2\rvert}\right)\right]^{-1} \lvert\mathcal{M}_{\gamma l_i\to \eta l_i}\rvert^2.
\ee
where the incoming momenta, $\vec{p}_1$ and $\vec{q}$, are co-linear while $\lvert \vec{p}_1\rvert\neq \lvert \vec{q}\rvert$. 
The scattering angle $\omega$ is defined to be the angle between $\vec{p}_1$ and $\vec{p}_2$. The momentum $\lvert\vec{p}_2\rvert$ is then given by
\be
\lvert\vec{p}_2\rvert=\frac{-\beta-\sqrt{\beta^2-4\alpha\gamma}}{2\alpha},
\ee
where
\bea
&&
\alpha=4\left[(\lvert\vec{p}_1\rvert-\lvert\vec{q}\rvert)^2\cos^2\omega-(E_1+E_q)^2\right]
\nonumber\\[1mm]
&&
\beta=4(\lvert\vec{p}_1\rvert-\lvert\vec{q}\rvert)\cos\omega\left[2m_i^2-m_\eta^2+2E_1 E_q +2\lvert\vec{p}_1\rvert \lvert\vec{q}\rvert\right]
\nonumber\\[1mm]
&&
\gamma=[m_i^2+2E_1 E_q+2\lvert\vec{p}_1\rvert \lvert\vec{q}\rvert]^2+(m_i^2-m_\eta^2)^2
\nonumber\\[1mm]
&&
\quad\quad+2m_\eta^2[-m_i^2-2\lvert\vec{p}_1\rvert \lvert\vec{q}\rvert-2 E_1 E_q]
\nonumber\\[1mm]
&&
\quad\quad -2m_i^2 [2E_1^2+2E_q^2-m_i^2+2E_1 E_q-2\lvert\vec{p}_1\rvert \lvert\vec{q}\rvert].
\eea

\section{The freeze-in production of $\eta$}
\label{freeze-in}
Consider the 2 to 2 annihilation process $AB\to XY$, where A and B are bath particles and the final state particles X and Y contain one or two $\eta$ particles. 
The Boltzmann equation for the freeze-in production of $\eta$ through this process is given by
\bea
&&\dot{n}_\eta+3Hn_\eta=-\int d\Pi_{X}d \Pi_Yd\Pi_A d\Pi_B (2\pi)^2\delta^{(4)}(p_X+p_Y-p_A-p_B)\times\\
&&\times\Bigg[\lvert\mathcal{M}_{X+Y\to A+B}\rvert^2 f_X f_Y(1\pm f_A)(1\pm f_B)
-\lvert\mathcal{M}_{A+B\to X+Y}\rvert^2 f_A f_B(1\pm f_X)(1\pm f_Y)\Bigg],
\nonumber
\eea  
which could be written as the following one dimensional integral
 \bea
 \dot{n}_\eta+3Hn_\eta
 &=&
 \frac{T}{32\pi^4}\int^\infty_{(m_A+m_B)^2}ds g_A g_B \left[s-(m_A+m_B)^2\right] \times
\nonumber\\
&&
\hspace{30mm}
\times\left[s-(m_A-m_B)^2\right]\frac{1}{\sqrt{s}}\sigma_{AB\to XY}(s)K_1\left(\frac{\sqrt{s}}{T}\right),
\eea  
where $K_1(x)$ is a modified Bessel function of second kind.
Most of the freeze-in production occurs at low temperatures where we approximate the annihilation cross section to be
\be
\sigma_{AB\to XY}(s)\approx \sigma_{AB\to XY}((m_A+m_B)^2).
\ee 
This will allow a further simplification of the Boltzmann equation,
\bea
\dot{n}_\eta+3Hn_\eta
&=&
\frac{T}{8\pi^4} g_A g_B (m_A m_B)^{3/2}\left[v_{rel}\sigma_{AB\to XY}(s=(m_A+m_B)^2)\right] \times
\nonumber\\
&&
\hspace{20mm}
\times
\int^\infty_{(m_A+m_B)^2}ds \sqrt{s-(m_A+m_B)^2}\frac{1}{\sqrt{s}}K_1\left(\frac{\sqrt{s}}{T}\right),
\eea
where $v_{rel}$ is the relative velocity of the incoming particles.  
The Boltzmann equation can now be written in terms of the yield, $y$,
\be
\dot{n}_\eta+3Hn_\eta=-1.66\frac{2\pi^2 g_{\ast s}\sqrt{g_\ast^\rho}T^6}{45 m_{PL}}\frac{dy}{dT},
\ee
where $g_{*}^\rho$ is is the number of relativistic degrees of freedom related to energy density. The yield is then calculated to be
\bea
y_{today}&=&
\frac{45 \;m_{PL} \;g_A g_B (m_A m_B)^{3/2}}{ 1.66 \times 16\pi^6}\left[v_{rel} \sigma_{AB\to XY}\left(s=(m_A+m_B)^2\right)\right]\times
\\
&&
\hspace{5mm}
\times\int^{T_{max}=\infty}_{T_{min}=0}
dT\frac{1}{g_{\ast s}(T)\sqrt{g^{\rho}_\ast(T)}T^5}\int^{\infty}_{(m_A+m_B)^2}ds \sqrt{s-(m_A+m_B)^2}\frac{1}{\sqrt{s}}K_1\left(\frac{\sqrt{s}}{T}\right).
\nonumber
\eea

\end{document}